\documentclass[journal=nalefd]{achemso}
\setkeys{acs}{articletitle=true}
\pdfoutput=1

\usepackage[version=3]{mhchem} 
\usepackage{amsmath}
\usepackage[colorlinks = true,
linkcolor = blue,
urlcolor = blue,
citecolor = blue,
anchorcolor = blue]{hyperref}
\usepackage{pifont}
\usepackage{ulem}
\usepackage{xcolor}
\usepackage{lineno} 

\usepackage{etoolbox} 
\makeatletter
\patchcmd{\acs@contact@details}{E}{*\,E}{}{}
\patchcmd{\acs@email@list@aux}{;}{\par*\,E-mail:}{}{} 
\makeatother

\newcommand{\Ang}{\text{\AA}}

\newcommand{\change}{\textcolor{black}} 
\newcommand{\degC}{$^\circ$C}
\newcommand{\Td}{T\textsubscript{d}}

\author{Chia-Hao Lee}
\affiliation{Department of Materials Science and Engineering, University of Illinois Urbana--Champaign, Urbana, Illinois 61801, United States}
\alsoaffiliation{These authors contributed equally to this work}

\author{Huije Ryu}
\affiliation{Department of Materials Science and Engineering, Seoul National University, Seoul 08826, Korea}
\alsoaffiliation{These authors contributed equally to this work}

\author{Gillian Nolan}
\affiliation{Department of Materials Science and Engineering, University of Illinois Urbana--Champaign, Urbana, Illinois 61801, United States}

\author{Yichao Zhang}
\affiliation{Department of Materials Science and Engineering, University of Illinois Urbana--Champaign, Urbana, Illinois 61801, United States}

\author{Yangjin Lee}
\affiliation{Department of Physics, Yonsei University, Seoul 03722, Korea}

\author{Siwon Oh}
\affiliation{Department of Physics, Sogang University, Seoul 04107, Korea}

\author{Hyeonsik Cheong}
\affiliation{Department of Physics, Sogang University, Seoul 04107, Korea}

\author{Kenji Watanabe}
\affiliation{Research Center for Functional Materials, National Institute for Materials Science, 1-1 Namiki, Tsukuba 305-0044, Japan}

\author{Takashi Taniguchi}
\affiliation{International Center for Materials Nanoarchitectonics, National Institute for Materials Science, 1-1 Namiki, Tsukuba 305-0044, Japan}

\author{Kwanpyo Kim}
\affiliation{Department of Physics, Yonsei University, Seoul 03722, Korea}

\author{Gwan-Hyoung Lee}
\email{gwanlee@snu.ac.kr}
\affiliation{Department of Materials Science and Engineering, Seoul National University, Seoul 08826, Korea}

\author{Pinshane Y. Huang}
\email{pyhuang@illinois.edu}
\affiliation{Department of Materials Science and Engineering, University of Illinois Urbana--Champaign, Urbana, Illinois 61801, United States}
\alsoaffiliation{Materials Research Laboratory, University of Illinois at Urbana--Champaign, Urbana, Illinois 61801, United States}

\title{\textit{In situ} Imaging of an Anisotropic Layer-by-Layer Phase Transition in Few-Layer \ce{MoTe2}}

\begin{document}
\newpage

\begin{abstract}
Understanding the phase transition mechanisms in two-dimensional (2D) materials is a key to precisely tailor their properties at the nanoscale. Molybdenum ditelluride (\ce{MoTe2}) exhibits multiple phases at room temperature, making it a promising candidate for phase-change applications. Here, we fabricate lateral 2H-\Td\ interfaces with laser irradiation and probe their phase transitions from micro- to atomic scales with \textit{in situ} heating in the transmission electron microscope (TEM). By encapsulating the \ce{MoTe2} with graphene protection layers, we create an \textit{in situ} reaction cell compatible with atomic resolution imaging. We find that the \Td-to-2H phase transition initiates at phase boundaries at low temperatures (200--225 \degC) and propagates anisotropically along the b-axis in a layer-by-layer fashion. We also demonstrate a fully reversible 2H-\Td-2H phase transition cycle, which generates a coherent 2H lattice containing inversion domain boundaries. Our results provide insights on fabricating 2D hetero-phase devices with atomically sharp and coherent interfaces. 

\end{abstract}

\subsection{Keywords}
\textit{In situ} heating, anisotropic phase transition, laser irradiation, molybdenum ditelluride, transmission electron microscopy, 2D materials

\leavevmode 

\begin{figure}[H]
\includegraphics[width= 16 cm]{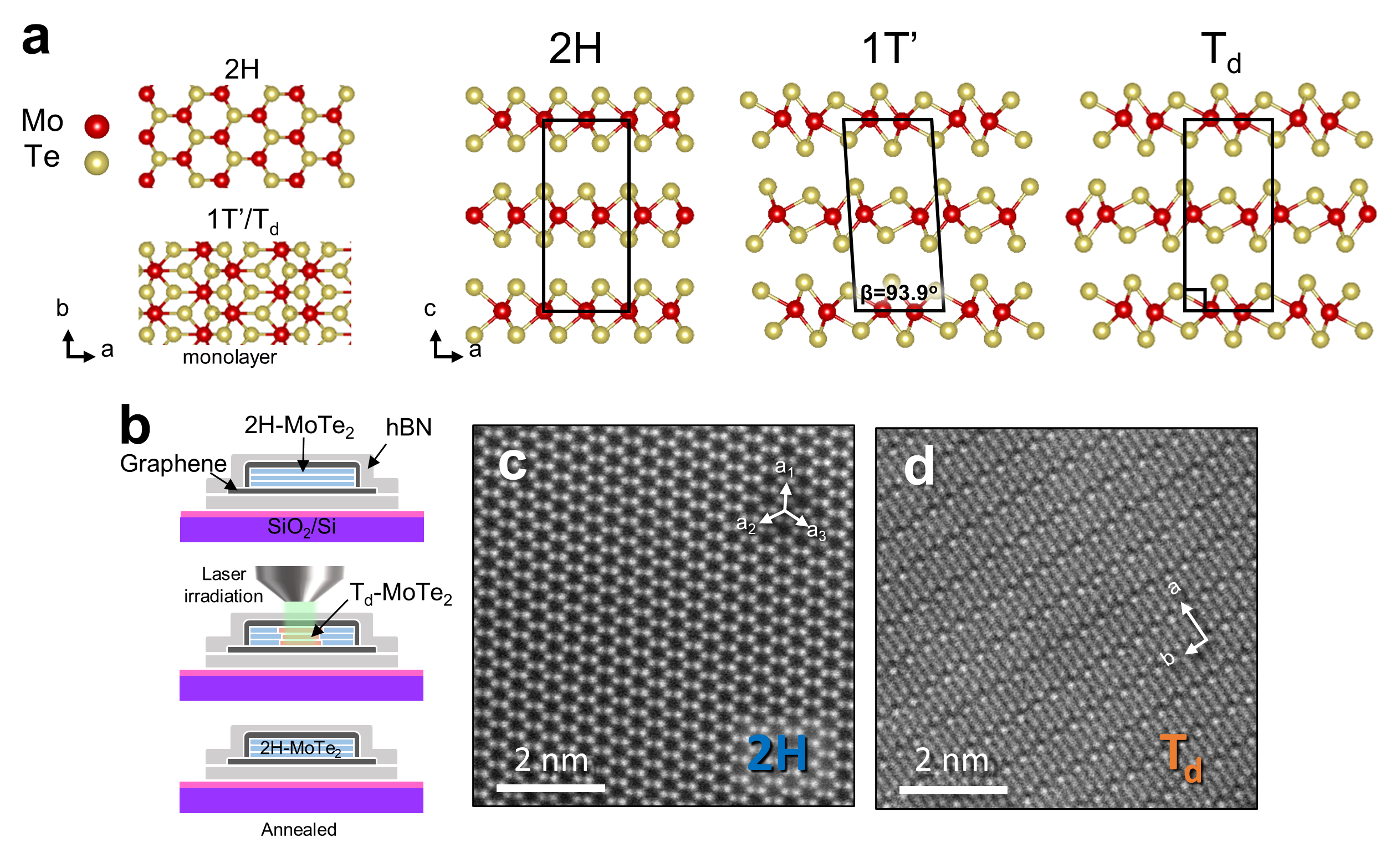}
\caption{Characterization and fabrication of different phases of \ce{MoTe2}. (a) Atomic structure models of 2H-, 1T'-, and \Td-\ce{MoTe2} with top and side views. Monolayer models are made for top view for clarity. (b) Schematic of the reversible phase transition of \ce{MoTe2}. The 2H-\ce{MoTe2} flakes are encapsulated by graphene and hBN layers. Local laser-irradiation induces 2H-to-\Td\ phase transition of \ce{MoTe2}, while the \Td\ phase reverts back to 2H phase after thermal annealing. (c,d) Aberration-corrected ADF-STEM images for the 2H and \Td\ phases, respectively.}
\label{fgr:1}
\end{figure}

Phase transformations in two-dimensional transition metal dichalcogenides (2D TMDCs) are an emerging research area due to their polymorphism—the ability to host different phases of the same chemical composition with distinct crystal structures. Utilizing phases with diverse electronic properties, multiple functionalities can be compactly packaged into nanoscale devices, such as monolithic 2D electronics \cite{Sung2017, Ma2019, Zhang2019e} and phase-change memories \cite{Zhang2019f, Zhu2019}. Among the group VI 2D TMDCs, molybdenum ditelluride (\ce{MoTe2}) has been widely studied because of the minimal energy difference (40 meV per formula unit) \cite{Duerloo2014, Duerloo2016, Kim2017} between the trigonal prismatic (2H), monoclinic (1T'), and orthorhombic (\Td) phases shown in Figure \ref{fgr:1}a. While the honeycomb lattice of the 2H structure is distinct, the 1T' and \Td\ phase share the same monolayer structure, but have different stacking structure in multilayers. The 1T' phase has a monoclinic structure with $\beta$ = $93.9^{\circ}$, while the \Td\ phase is orthorhombic with $\beta$ = $90^{\circ}$ \cite{Manolikas1979}. This stacking difference results in the broken inversion symmetry of the \Td\ phase and its unique quantum properties, including type-II Weyl fermions\cite{Sun2015, Soluyanov2015}, quantum spin Hall effect\cite{Song2020}, giant magnetoresistance\cite{Lee2018}, and superconductivity\cite{Qi2016}. 

Phase transitions between 2H- and 1T'-\ce{MoTe2} have been demonstrated using electric biasing \cite{Wang2017, Zhang2019f}, strain\cite{Song2016}, heat\cite{Keum2015, Wang2022}, ion intercalation\cite{Eshete2019}, and laser irradiation\cite{Cho2015, Tan2018, Bae2021}. However, the phase transitions between 2H- and \Td-\ce{MoTe2} remain largely unexplored because the \Td\ phase is less thermodynamically stable than other phases under ambient conditions, which makes it difficult to study its room temperature properties and phase-change behaviors. Conventionally, the \Td\ phase is obtained by cooling 1T'-\ce{MoTe2} crystals down to 250 K\cite{Zhang2016, Chen2016} or through chemically alloying with W substitutions\cite{Rhodes2017, Lv2017, Jin2018}. Very recently, the 2H-to-\Td\ transition has been reported with high temperature annealing of hBN-encapsulated \ce{MoTe2} \cite{Ryu2021}, \change{while this work focuses on the reverse \Td-to-2H transition and its atomic-scale mechanisms.} Understanding the micro- to atomic scale phase transition mechanisms between the semiconducting 2H and the topological \Td\ phase may open up new possibilities towards low-dissipation 2D electronics and spintronics.

\textit{In situ} TEM is a powerful technique to investigate phase transformations in 2D materials\cite{Lin2014, Ryu2019}. However, electron beam irradiation\cite{Elibol2018, Lehnert2019} and vacuum annealing \cite{Zhu2017a, Zhu2017} can cause major degradation of \ce{MoTe2} and other 2D TMDCs due to the significant loss of chalcogen atoms, making it particularly challenging to probe their phase transitions without altering the chemical composition. 

Here, we combine \textit{in situ} TEM with graphene encapsulation to study the reversible phase transitions of \ce{MoTe2} from micro- to atomic scales. We first use laser irradiation to locally convert few-layer \ce{MoTe2} flakes from the 2H to a mixture of 1T' and \Td\ phases, which we find is primarily \Td\ in the regions examined by our TEM experiments. Then, we apply \textit{in situ} pulsed heating to monitor the reverse phase transition from the \Td\ to 2H phase with a combination of aberration-corrected scanning transmission electron microscopy (STEM) and dark-field TEM (DFTEM). We find that the \Td-to-2H phase transition initiates at the 2H-\Td\ interface at around 200--225 \degC. Between 200--400 \degC, we observe a highly anisotropic phase transition: the 2H phase fronts progress along the b-axis of the \Td\ grains, in a layer-by-layer fashion. The ability to visualize each 2H phase front enables measurements of \Td-to-2H phase transition kinetics of individual \ce{MoTe2} layers. Lastly, we demonstrate the reversibility of phase transitions between 2H and \Td\ phases with cycles of laser irradiation and vacuum heating.


Figure \ref{fgr:1}b shows a schematic of the phase conversion process. To create an encapsulated cell, we fabricate hBN/graphene/\ce{MoTe2}/graphene/hBN heterostructures using a PDMS-assisted pick-up technique\cite{Wang2013}. The \ce{MoTe2} flakes are mechanically exfoliated with lateral size around tens of microns and 4--5 layers in thickness. The \ce{MoTe2} is encapsulated by both monolayer graphene and 10 nm thick hBN on the top and bottom surfaces. The hBN layers improve adhesion with the polymer film used in the pick-up technique and are removed before (S)TEM analysis using \ce{XeF2} etching\cite{Son2018} (see Supporting Information (SI) \change{Section 1 and} Figure S1). The encapsulation is essential because it creates an enclosed reaction cell that acts as physical and chemical barrier for \ce{MoTe2}, minimizing the sublimation of Te and interactions with the atmosphere during further processing. \change{If the \ce{MoTe2} were not encapsulated, it would be nearly impossible to observe the phase transition without modifying the crystal stoichiometry through the loss of Te atoms, which has been shown to impact the phase transition. Using encapsulated samples, we did not observe any Te vacancy formation via ADF-STEM during heating.} Importantly, the graphene contributes minimal background signal to the TEM images, enabling atomic-resolution imaging\cite{Huang2012, Algara-Siller2013, Elibol2018}. 

We then irradiate the encapsulated 2H-\ce{MoTe2} with a 532 nm laser to locally initiate the phase transition from the 2H to a primarily \Td\ phase \change{(SI Section 2)}. \change{Because it is difficult to distinguish 1T' and \Td\ phases, the phases of pristine and laser-irradiated \ce{MoTe2} are characterized by multiple techniques, including aberration-corrected annular dark-field STEM (ADF-STEM) images (Figure \ref{fgr:1}c--d), TEM diffraction, and polarized Raman spectroscopy (SI Figure S2).} TEM diffraction and polarized Raman measurements indicate that the resulting materials contain a mixture of 1T' and \Td\ phases, which is in agreement with previous reports \cite{Huang2019a, Hart2022}. The potential for mixed phases occurs because the calculated energy difference between 1T' and \Td\ phase is less than 3 meV per unit cell \cite{Kim2017, Huang2019a}. In the TEM samples analyzed below, however, atomic resolution STEM imaging (Figure \ref{fgr:1}d) indicates that the laser-irradiated material is primarily \Td\ (see SI Figure S3 for top-down ADF-STEM image simulation of 1T' and \Td\ phases). Therefore, we refer to the transformed phase as \Td.

\begin{figure}[H]
\includegraphics[width= 16 cm]{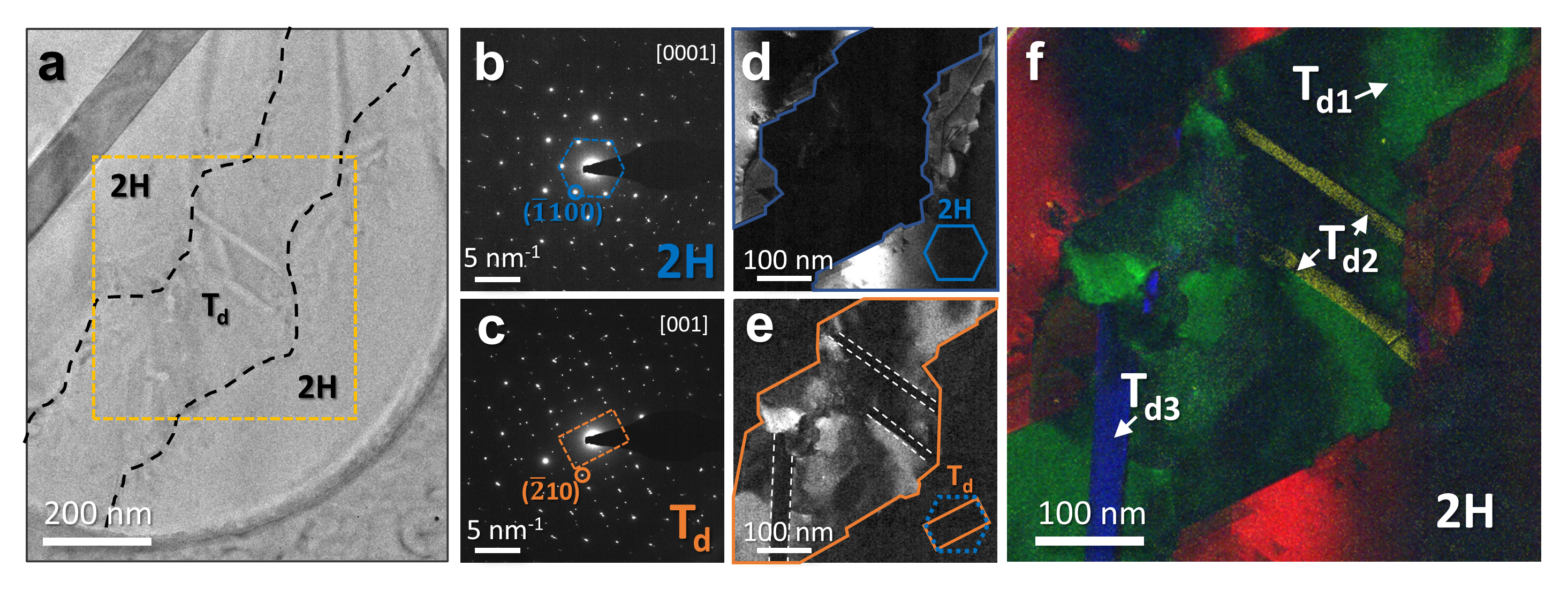}
\caption{Phase and grain orientation mapping of laser-irradiated \ce{MoTe2} with DFTEM. (a) BFTEM image of the suspended, graphene-encapsulated \ce{MoTe2} containing both 2H and \Td\ grains. The \Td\ phase region is delineated by the laser trajectory and outlined by the black dashed lines. \change{The minimum width of the \Td\ region is determined by the radial laser intensity profile.} (b,c) SAED patterns, (d,e) DFTEM images of 2H and \Td\ phase, respectively. The diffraction patterns (b,c) are acquired with zone axis perpendicular to the basal planes. The weaker diffraction spots are generated by the graphene encapsulating layers. The DFTEM images (d-e) are formed by selecting the $(\bar{1}100)$\textsubscript{2H} and $(\bar{2}10)$\textsubscript{\Td} Bragg reflections in (b) and (c) with the objective aperture position marked with blue and orange circles. The objective aperture and selected Bragg reflections are centered on the optical axis to reduce image aberrations. (f) Overlay of false-colored DFTEM images of the 2H matrix (red) and three different orientations of \Td\ grains (green, yellow, and blue). The DFTEM images (d--f) are acquired at the region marked by the yellow dashed square in (a).} 
\label{fgr:2}
\end{figure}

For \textit{in situ} heating, we transfer the laser-irradiated, encapsulated \ce{MoTe2} specimens to a microelectromechanical system (MEMS)-based heating TEM chip \change{(SI Section 1)}. Bright-field TEM (BFTEM) imaging before \textit{in situ} heating (Figure \ref{fgr:2}a) shows very little contrast between the 2H and \Td\ phases, indicating a uniform thickness across the hetero-phase interface. The selected-area electron diffraction (SAED) patterns in Figure \ref{fgr:2}b--c exhibit the characteristic hexagonal and rectangular lattice of the 2H and \Td\ phases. 

We use DFTEM to map the real-space location and orientation of the 2H and \Td\ phases \change{(SI Section 3)}. DFTEM has been widely used to determine the crystal orientation and stacking order of 2D materials\cite{Huang2011, Ping2012, VanDerZande2013} and operates by selecting specific Bragg spots in the diffraction pattern with an objective aperture, so that only crystal grains that diffract to a narrow range of k-vectors appear bright in the image. DFTEM images of the 2H and \Td\ phases in Figure \ref{fgr:2}d--e are obtained by selecting the $(\bar{1}100)$\textsubscript{2H} and $(\bar{2}10)$\textsubscript{\Td} Bragg reflections, marked with blue and orange circles. We observe \Td\ grains with three orientation directions, rotated 120$^\circ$ from each other. The b-axis of each \Td\ orientation is parallel to one of the three zig-zag directions of the three-fold symmetric 2H matrix (SI Figure S4). Figure \ref{fgr:2}f shows a false-colored DFTEM overlay image mapping the four grains present after laser-irradiation: the 2H phase (red) and the three \Td\ orientations (green, yellow, and blue). The majority of the \Td\ region in Figure \ref{fgr:2}f is oriented in one of the three orientations (green), with needle-like inclusions of the other two orientations. 


Next, we perform \textit{in situ} heating to investigate the reverse \Td-to-2H phase transition. We use heat pulses\cite{Wang2022} instead of continuous heating for three reasons: (1) Pulsing provides flexibility to ``halt'' the phase transition at any time, rapidly jump to specific temperatures, and even hold at different temperatures for more detailed kinetic studies. (2) The ability to pause between pulses makes it possible to acquire both large field-of-view (FOV) DFTEM and atomic-resolution ADF-STEM images at several positions between pulses, which provides both a large-scale view of the phase transition kinetics and atomic scale snapshots at the interfaces. (3) Heat pulsing minimizes the energy input and potential sublimation during the phase transition. While the graphene encapsulation minimizes damage and sublimation to the \ce{MoTe2}, we find that heating at temperatures above 600 \degC\ for 0.5 s can produce small (5--10 nm) voids (see SI Movie S1). 

\begin{figure}[H]
\includegraphics[width= 16 cm]{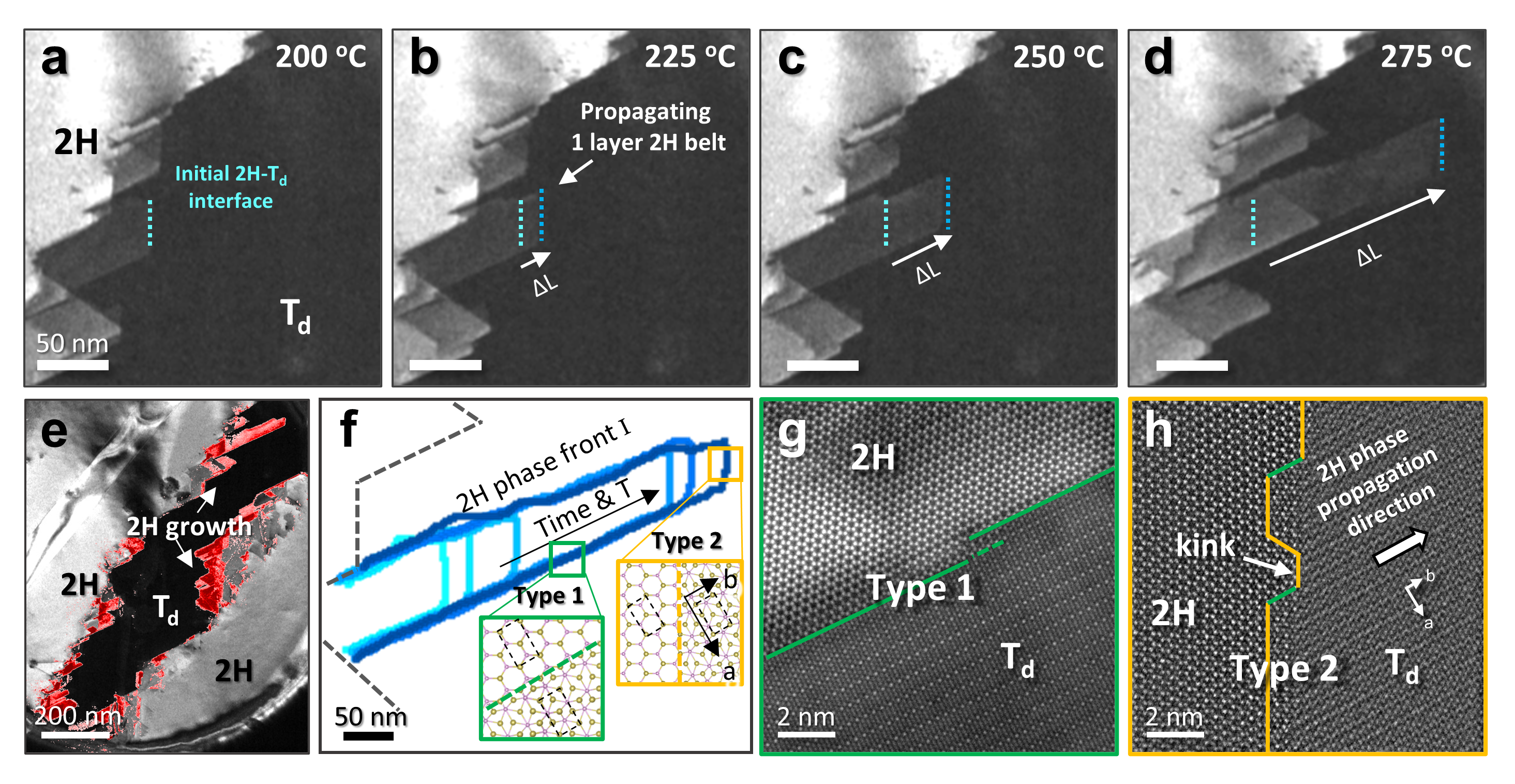}
\caption{Anisotropic, low-temperature \Td-to-2H phase transition. (a--d) DFTEM images formed from the $(\bar{1}100)$\textsubscript{2H} spot are acquired at room temperature after heat pulses of 0.5 s from 200 to 275 \degC. The \Td-to-2H transition initiates at the interface, and the 2H phase front anisotropically propagates into the \Td\ phase region. (e) Overlay of a low magnification DFTEM image with newly grown 2H regions marked in red. The \Td-to-2H phase transition occurs primarily at 2H-\Td\ interfaces. (f) Contour plot of the 2H phase front in the same region as (a--d) shows propagation along the b-axis of nearby \Td\ grain. The insets are the atomic models of 2 different types of interface. The anisotropy arises from the different interface energy of type 1 and 2 interfaces. ADF-STEM images of (g) type 1 and (h) type 2 2H-\Td\ interfaces with atomic kinks indicates a step-flow growth model.}
\label{fgr:3}
\end{figure}

\begin{figure}[H]
\includegraphics[width= 16 cm]{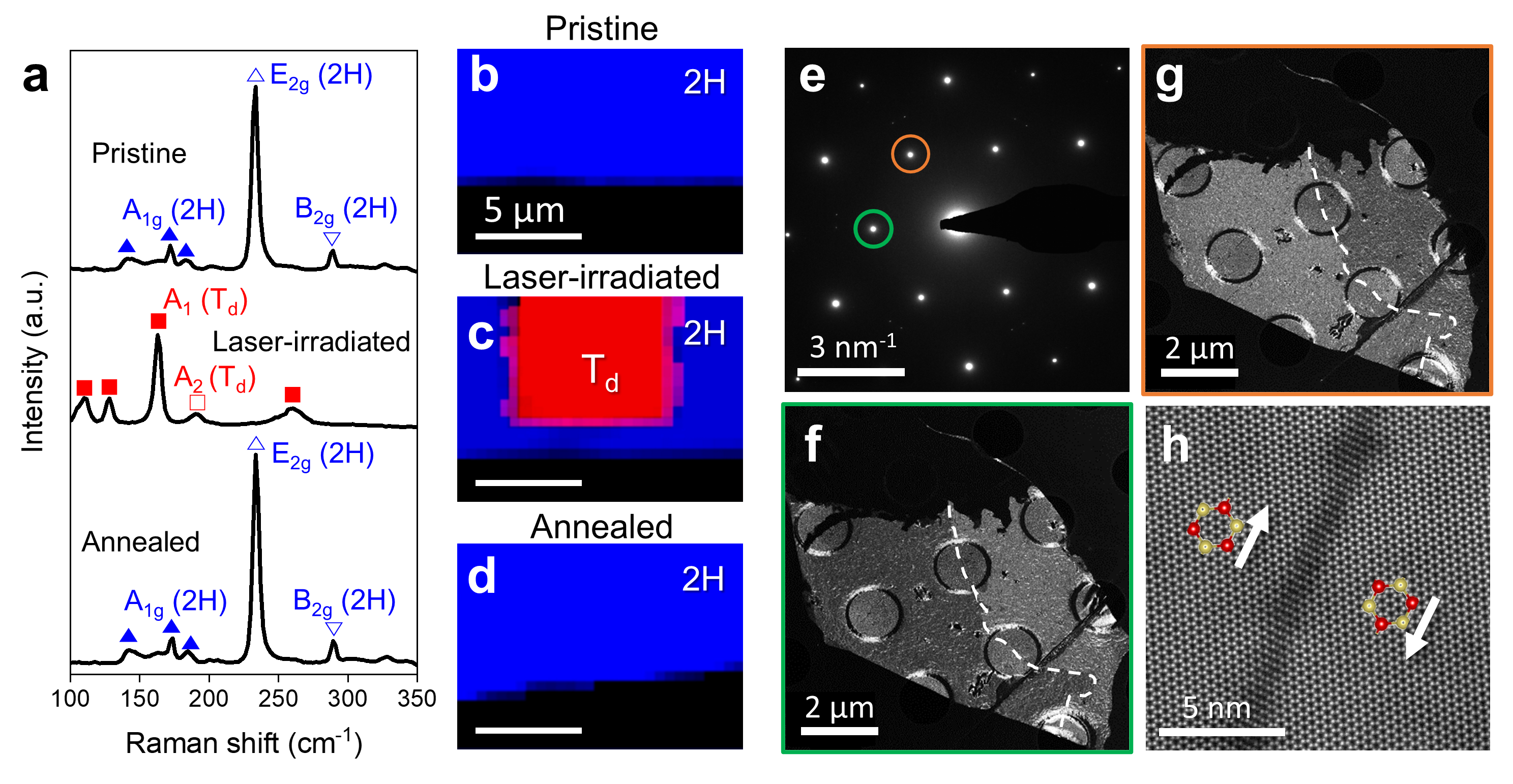}
\caption{Cyclic phase transition and recovery of \ce{MoTe2} (2H $\xrightarrow{}$ \Td\ $\xrightarrow{}$ 2H phase) via laser irradiation and annealing. (a) Raman spectra and (b--d) Raman mapping of pristine (as-stacked), laser-irradiated, and annealed \ce{MoTe2}. The Raman maps are visualized with the E\textsubscript{2g} (2H) and A\textsubscript{1} (\Td) peak, respectively. (e) SAED pattern of the 2H phase region. The orange and green circles denote the objective aperture positions that are used to generate DFTEM images (f,g) from different $(\bar{1}100)$\textsubscript{2H} Bragg reflections. The brighter region corresponds to the specific 2H orientation that generates the stronger Bragg reflection. The inversion domain boundary is outlined by the white dashed line. (h) ADF-STEM image at the inversion domain boundary of 2H grains with opposite orientations. The atomic models are overlaid with arrows indicating opposing orientations.}
\label{fgr:5}
\end{figure}

We use DFTEM imaging to track the propagation of 2H phase between heat pulses, with pulse durations of 0.5 to 60 s, and temperatures from 200 to 275 \degC; note that all images are acquired between pulses, when the sample is at room temperature. DFTEM images of the 2H phase after different heat pulse temperatures (Figure \ref{fgr:3}a--d and SI Movie S1 ) show that the 2H region at the phase boundary propagates anisotropically toward the \Td\ grain during heating, forming a belt-shaped inclusion. Figure \ref{fgr:3}e shows a large-FOV DFTEM image where newly grown 2H regions, marked in red, inherit the orientation of the 2H matrix. We occasionally observed inversion domains in the 2H phase, which we discuss later in Figure \ref{fgr:5}. Contrary to previous reports of high (500--600 \degC) 1T'-to-2H phase transition temperatures in bulk samples\cite{Xu2019a}, we observe the \Td-to-2H phase transition initiates at temperatures as low as 200--225 \degC, from the existing 2H-\Td\ interfaces. \change{There are two reasons for such low transition temperatures: First, the \Td\ phase is thermodynamically unstable under ambient condition, so only the kinetic barrier needs to be overcome. Second, the existing 2H-\Td\ interfaces act as nucleation sites, which further reduce the kinetic barrier of the \Td-to-2H phase transition.}

Figure \ref{fgr:3}f shows a contour overlay of the 2H phase fronts captured between heat pulses of 200--275 \degC\ in a 4-layer thick sample. We outline the propagating 2H regions that contain at least a monolayer of 2H phase. This image shows that the 2H phase growth is anisotropic in-plane, progressing along the [010]\textsubscript{\Td} (b-axis direction) of the \Td\ grain. This result is in contrast to previous reports of an isotropic 1T'-to-2H transition, which is an averaged result from large-scale polycrystalline 1T' grains\cite{Xu2019a, Xu2021}. The preferential b-axis growth of the 2H phase is observed for all \Td\ orientations (SI Figure S5). The anisotropy occurs mainly at low-temperatures, and we find the phase transition becomes more isotropic above 400 \degC\ (SI Movie S2). 

As shown in the atomic models in Figure \ref{fgr:3}f, 2H-\Td\ phase boundaries can be classified into two types\cite{Sung2017} based on their symmetry: Type 1, where the phase boundary is parallel to the b-axis of \Td\, and Type 2, where the phase boundary is rotated by 120$^\circ$ from the b-axis of \Td. The anisotropic propagation of the 2H phase suggests that the propagation (growth) rate of the type 2 interface is much faster than for type 1, resulting in the formation of belt-shaped 2H grains with mostly type 1 interfaces. This behavior can be described by a kinetic Wulff construction\cite{Artyukhov2012, Ma2013, Zhao2017a, Lin2021}, where the final crystal shape is predicted using thermodynamic and kinetic factors including the interface energy and relative growth rate of different facets. The type 2 interface energy is estimated to be 70 meV/\Ang\ higher than type 1 interface\cite{Berry2018}, making type 1 interfaces more thermodynamically stable. This is consistent with our observations that there are more kinks and steps in the atomic-resolution ADF-STEM images of type 2 interfaces than in the type 1 interfaces (Figure \ref{fgr:3}g,h). The 2H growth preferentially propagates along the [010]\textsubscript{\Td} direction due to the higher kink formation and expansion rate of type 2 interfaces.

\begin{figure}[H]
\includegraphics[width= 16 cm]{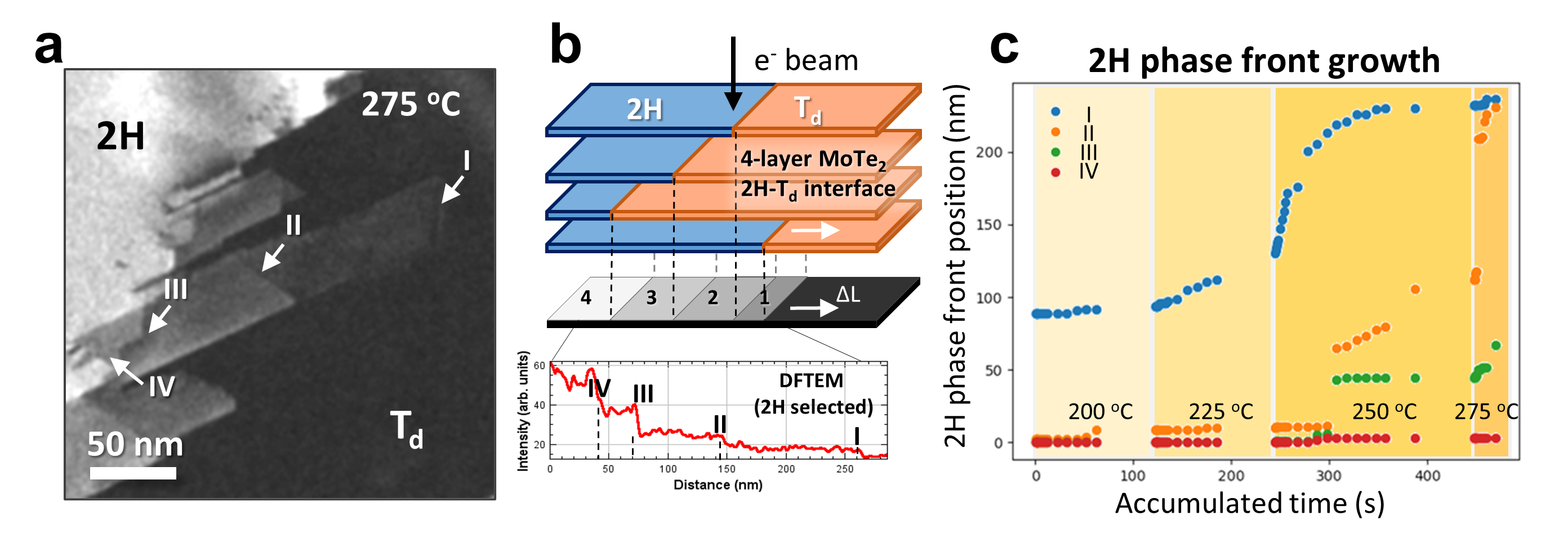}
\caption{Layer-by-layer phase transition and growth kinetics measurement. (a) DFTEM image that shows the layer-by-layer phase transition. The 2H-\Td\ interface of different layers are individually identified by their intensity difference. (b) Schematic of the intensity differences of 4-layer \ce{MoTe2} 2H-\Td\ interfaces in DFTEM. The mono-, bi-, tri-, and quad-layer 2H phase fronts are labeled as I, II, III, and IV respectively. Note that the relative positions (in the z-direction) of each phase front are unknown due to the projection nature of TEM. (c) Plot of 2H phase front positions of different layers as a function of accumulated heating time. We perform a series of short heat pulses to capture the phase transition. Each dot corresponds to a heat pulse. The pulsing time ranges from 0.5 s to 1 min and can be read from the horizontal spacing between the dots. The pulsing temperatures (200--275 \degC) are color-coded by the background shades. The propagation rates are extracted by the slope of the curves, which have a strong temperature dependence.}
\label{fgr:4}
\end{figure}

Figure \ref{fgr:4}a shows that the newly grown 2H region in the 4-layer \ce{MoTe2} has multiple phase fronts ($I$, $II$, $III$, and $IV$). Figure \ref{fgr:4}b schematically shows horizontally staggered 2H phase fronts and the resulting DFTEM intensities. In the kinematic limit, DFTEM intensities scale quadratically with the number of 2H layers, making it possible to individually probe the position of the 2H phase front at each layer. \change{However, we are not able to determine the depth of each phase front because TEM produces images that are averaged in projection (along the direction of the electron beam path).} In Figure \ref{fgr:4}c, we measure the 2H-\Td\ interface positions of each layer as a function of accumulated heating time. The calculated propagation rates range from 0.07 to 0.4 nm/s at 225 \degC\ and exhibit wide variability. For example, both interfaces $II$ (orange) and $III$ (green) exhibit a sudden jump in 2H phase front position at t = 300 sec after the fifteenth 250 \degC\ pulse is applied. The non-uniform propagation rates might be due to strain, defects, and differing surface energies between atomic layers (2H, \Td, and graphene encapsulation), which can locally alter the energy barrier of \ce{MoTe2} phase transitions\cite{Tang2018}. 

Next, we demonstrate that the laser-induced \Td\ phase can be transformed back to the 2H phase via \textit{ex situ} vacuum annealing (Figure \ref{fgr:5} \change{and SI Section 4}). We characterize the pristine, laser-irradiated, and annealed \ce{MoTe2} with Raman spectroscopy (Figure \ref{fgr:5}a--d \change{and SI Section 5}). Pristine (as-stacked) \ce{MoTe2} (Figure \ref{fgr:5}a,b) exhibits the three characteristic Raman peaks (E\textsubscript{2g}, A\textsubscript{1g}, and B\textsubscript{2g}) of the 2H phase, while laser-irradiated regions (red areas in Fig \ref{fgr:5}c) exhibit the A\textsubscript{1} and A\textsubscript{2} modes of the \Td\ phase. The Raman maps (Figure \ref{fgr:5}b,c) show that the \Td\ region transformed from 2H phase is uniform, with 2H-\Td\ boundaries that are sharp on the micron scale. After annealing at 800 \degC\ for 3 hours, Raman mapping indicates the structure is fully and uniformly converted to 2H phase, as shown in Figure \ref{fgr:5}d. This result shows that a reversible phase transition of 2H- and \Td-\ce{MoTe2} can be achieved by laser irradiation and vacuum annealing. When we anneal multiple samples at different temperatures (300--800 \degC), all of them exhibit the \Td-to-2H phase transition.

Finally, we examine the crystal structure of \ce{MoTe2} after a full conversion cycle from 2H, to \Td, and back to 2H phase in Figure \ref{fgr:5}e--h. The six-fold symmetry of the SAED pattern in Figure \ref{fgr:5}e indicates that the converted sample has no rotational grain boundaries. However, the DFTEM image (Figure \ref{fgr:5}f) shows that for a selected $(\bar{1}100)$\textsubscript{2H} Bragg reflection (marked with a green circle in Figure \ref{fgr:5}e), one of the two grains appears brighter due to the breaking of Friedel's law \cite{VanDerZande2013, Deb2020}. By selecting a neighboring Bragg reflection (orange circle in Figure \ref{fgr:5}f), the contrast of the two grains is reversed (Figure \ref{fgr:5}g). This indicates that the 2H grains have inversion symmetric orientations separated by an inversion domain boundary (IDB), a twin boundary commonly observed in 2D TMDCs \cite{VanDerZande2013, Zhu2017, Hong2017}. Figure \ref{fgr:5}h shows an atomic resolution STEM image of an IDB in the 2H phase region after a full conversion cycle of 2H-\Td-2H. We also observe an IDB running through only 3 of the layers in a 5-layer \ce{MoTe2} sample (SI Figure S6), indicating the IDBs do not necessarily go all the way through the sample. IDBs are likely generated during the \Td-to-2H phase transition because the \Td\ grain has two equivalent transition pathways (SI Figure S7). As a result, cyclic phase transitions from 2H-\Td-2H convert a 2D single crystal to coherent 2H polycrystals stitched with IDBs. Our work shows that cyclic phase transitions are a promising technique to fabricate the IDBs, which act as one-dimensional metallic tunnels\cite{Liu2014, Lehtinen2015} embedded in 2D semiconductors. 

In conclusion, we have demonstrated that encapsulated, few-layer \ce{MoTe2} can be reversibly phase engineered between the semiconducting 2H phase and the \Td\ phase using laser irradiation and thermal annealing. Using \textit{in situ} pulsed heating and DFTEM, we show that the \Td-to-2H phase transition initiates at the 2H-\Td\ interfaces at temperatures as low as 200--225 \degC. Moreover, we observe anisotropic growth of the 2H phase front, which preferentially propagates along the b-axis of the nearby \Td\ grains. Our findings can be applied to fabrication of coplanar 2D circuitry, including 2D Josephson junctions\cite{Madsen2017}, broadband photodetectors\cite{Lai2018}, and other hetero-phase devices. Finally, we demonstrate a new approach for \textit{in situ} studies of 2D materials using graphene encapsulation and pulsed heating, which can be applied to other micro- to atomic scale \textit{in situ} studies of solid state phase transitions.

\newpage

\subsection{\textbf{\ding{110} ASSOCIATED CONTENT}}
\textbf{Supporting Information}
Sample fabrication workflow, 1T' and \Td\ mixture analysis, simulated ADF-STEM images, ADF-STEM images and atomic models of inversion domain boundaries, and \textit{in situ} movies of \ce{MoTe2} phase transition.

\subsection{\textbf{\ding{110} AUTHOR INFORMATION}}
\noindent\textbf{Corresponding Author}\\
*Email: \href{pyhuang@illinois.edu}{pyhuang@illinois.edu}\\
*Email: \href{gwanlee@snu.ac.kr}{gwanlee@snu.ac.kr}

\noindent\textbf{Author Contributions}\\
Under supervision by P.Y.H., C.-H.L., G.N. and Y.Z. acquired and analyzed the \textit{in situ} heating DFTEM and ADF-STEM images. Under supervision by G.-H.L., H.R. fabricated the \ce{MoTe2} samples, performed \textit{ex situ} annealing experiments and Raman spectroscopy. Under supervision by K.K., Y.L. perform TEM analysis of phase-engineered \ce{MoTe2}. Under supervision by H.C., S.O. conducted polarized Raman measurements. K.W. and T.T. synthesized the hBN flakes. All authors read and contributed to the manuscript.

\noindent\textbf{Notes}\\
The authors declare no competing financial interest.

\begin{acknowledgement}
This material is based upon work supported by the U.S. Department of Energy, Office of Science, Office of Basic Energy Sciences, Division of Materials Sciences and Engineering under award number DE-SC0020190, which supported the electron microscopy and related data analysis. This work was carried out in part in the Materials Research Laboratory Central Facilities at the University of Illinois at Urbana--Champaign. G.-H.L. acknowledges support by the Creative-Pioneering Researchers Program through Seoul National University (SNU), the National Research Foundation (NRF) of Korea (NRF-2021R1A2C3014316, SRC program: vdWMRC center 2017R1A5A1014862, NRF-2021M3F3A2A01037858), the Research Institute of Advanced Materials (RIAM), Institute of Engineering Research, and Institute of Applied Physics at SNU, which supported the sample fabrication and ex situ characterization. K.W. and T.T. acknowledge support from the Japan Society for the Promotion of Science (JSPS) KAKENHI (Grant Numbers 19H05790 and 20H00354) and A3 Foresight by JSPS, which supported the h-BN synthesis.

\end{acknowledgement}

\bibliography{My-Collection}

\newpage
\subsection{Graphical TOC Entry}
\begin{figure}[H]
\includegraphics[width= 7.16 cm]{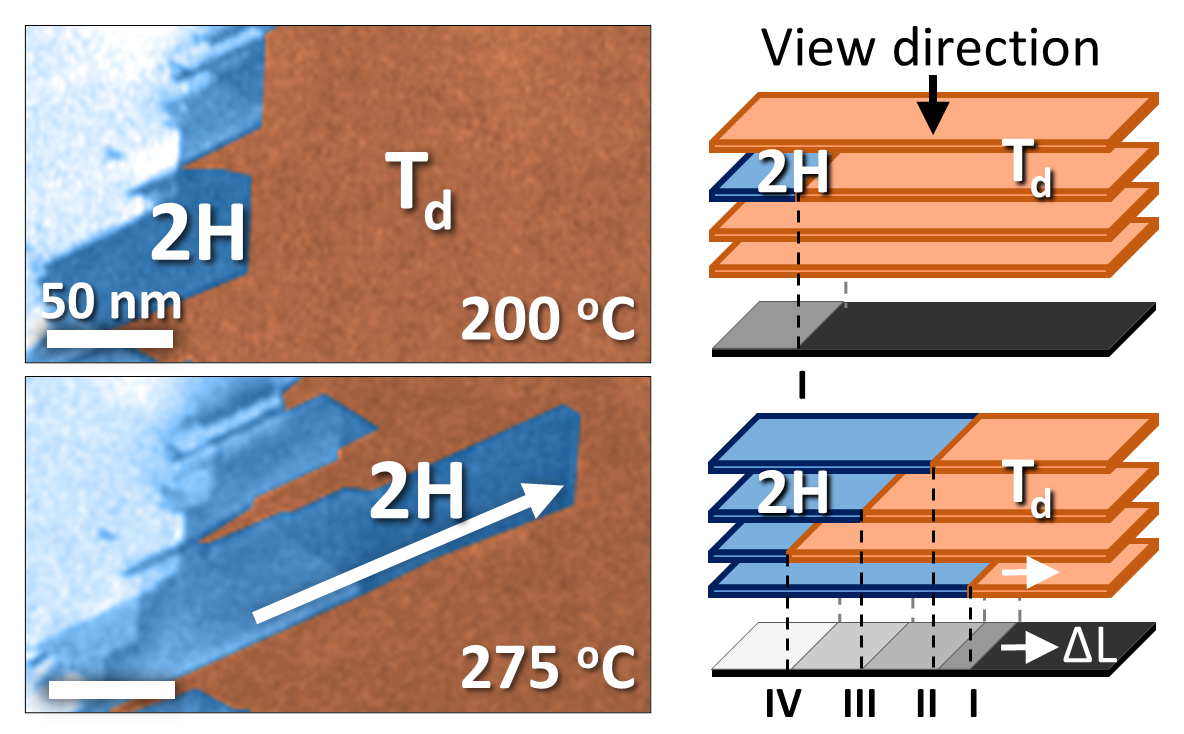}
\label{TOC}
\end{figure}

\end{document}


\newpage

\section{1. Sample preparation for \textit{in} and \textit{ex situ} experiments}

To fabricate the hBN/Gr/\ce{MoTe2}/Gr/hBN heterostructures, we mechanically exfoliated thin layers of 2D materials (\ce{MoTe2}, hBN, graphene) from bulk crystals (\ce{MoTe2}: HQ graphene, graphene: NGS Naturgraphite GmbH, hBN: NIMS) onto \ce{SiO2}/Si substrate. We then used the pick-up transfer technique\cite{Purdie2018} with a poly(bisphenol A carbonate, Sigma Aldrich) (PC)-coated poly (dimethyl siloxane) (PDMS) lens mounted on a microscope slide to pick-up and released the constituent flakes on the substrate. The PC/PDMS/glass slide was held in a 3-axis micromanipulator to control the position of the contact area with the 2D materials. The substrate was placed on a heating stage. By controlling the temperature of the heating stage (80--130 \degC), the 2D flakes were picked up by the PC with minimal cracking or folding, leaving the substrate on the heating stage. The hBN/Gr/\ce{MoTe2}/Gr/hBN heterostructures were fabricated by repeating the above steps, and then transferred onto a clean \ce{SiO2}/Si substrate by releasing the PC film from the PDMS lens at a temperature above 180 \degC. Lastly, we placed the entire sample in chloroform for 30 min to remove the PC film. 

\begin{figure}[H]
  \includegraphics[width=16cm]{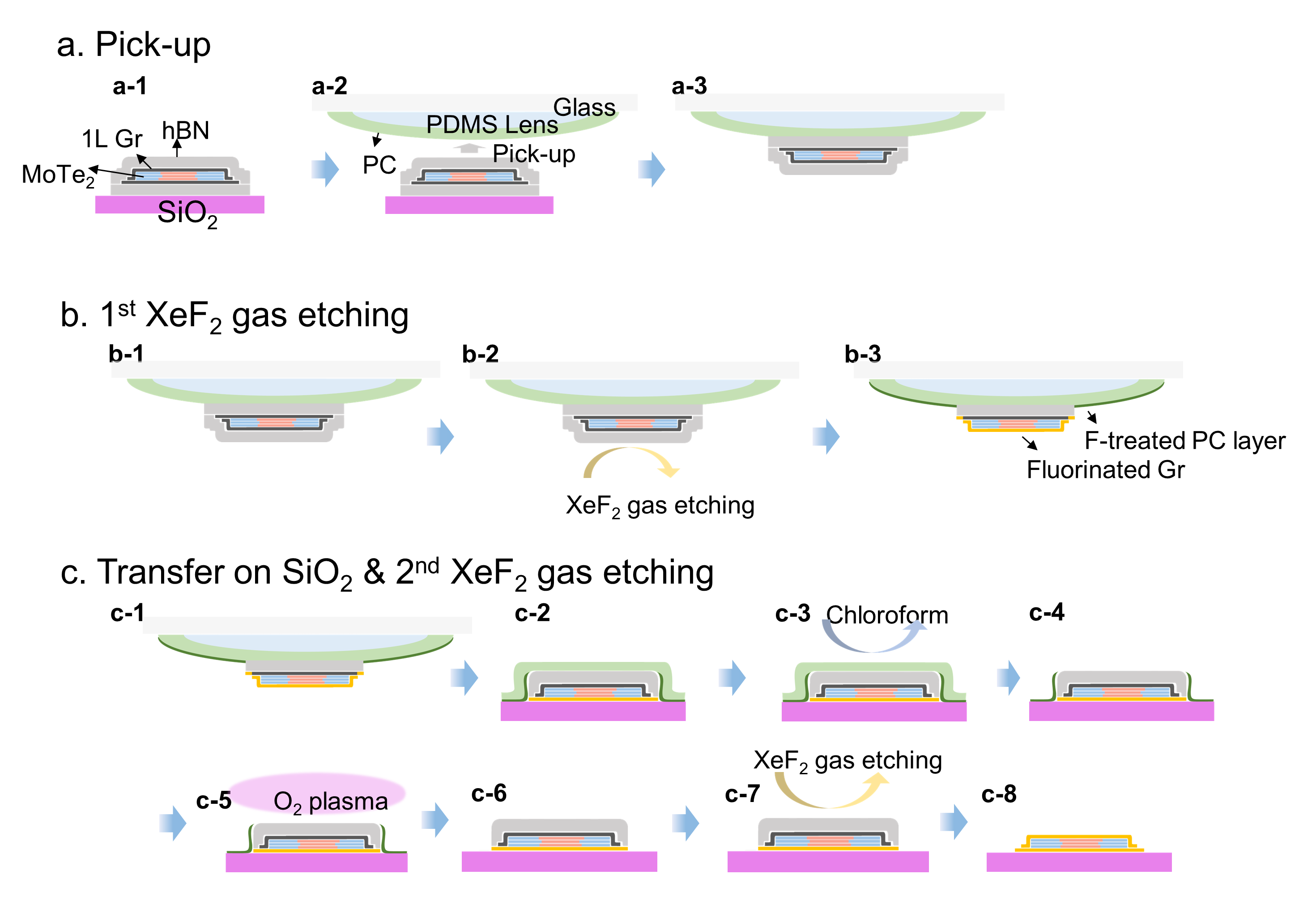}
  \caption{Sample preparation for \textit{in situ} TEM experiments. (a-1) The schematic illustration of hBN/Gr/\ce{MoTe2}/Gr/hBN structure on \ce{SiO2} (285 nm)/Si++ substrate. (a-2,3) Pick-up the structure by polycarbonate (PC) film on polydimethylsiloxane (PDMS). (b-1,2) Etching the bottom side of the hBN by exposing to \ce{XeF2} gas. (b-3) After exposing to \ce{XeF2} gas, the bottom hBN is completely etched, while the graphene layers and the encapsulated layers remain. In addition, the PC exposed by \ce{XeF2} is also chemically modified, which can not be dissolved by chloroform. (c-1,2) One-side-etched sample is transferred to another \ce{SiO2}(285 nm)/Si++ substrate at about 180 \degC\ and separated from the PDMS lens. The Si substrate was treated with \ce{O2} plasma to increase the adhesion energy of \ce{SiO2}. (c-3,4) Remove the PC film by chloroform bath. Note that the fluorinated PC was not dissolved in chloroform. (c-5,6) Etch off the fluorinated PC layer by \ce{O2} plasma. Since the etch rate of hBN is much slower than fluorinated PC layer using \ce{O2} plasma, the fluorinated PC layer is removed while the Gr/\ce{MoTe2}/fluorinated Gr structure remains. (c-7,8) Remove the top hBN by \ce{XeF2} gas etching. Finally, we transferred the heterostructures on MEMS TEM chips using the PMMA-assisted, wet-transfer method.} 
  \label{fgr:S1}
\end{figure}

To image the \ce{MoTe2} at atomic resolution and reduce multiple scattering from thick hBN layers, we removed the hBN layers with \ce{XeF2} dry etching\cite{Son2018}. The sample fabrication process was similar with the one used for \textit{ex situ} experiments but with some modifications, see Figure \ref{fgr:S1} for the schematic. The bottom hBN layer was etched away by the \ce{XeF2} exposure, and the etching process was self-limited at the graphene layer. We transferred the stack onto a clean \ce{SiO2}/Si substrate and exposed it with chloroform, oxygen plasma, and \ce{XeF2} again to remove the top hBN layers. After these steps, the stack was encapsulated with fluorinated graphene and ready for transferring onto an in situ heating TEM chip (E-FHDC-VO-10, Protochips). We used the conventional polymer transfer technique with poly(methyl methacrylate) (PMMA) and KOH to transfer the stack\cite{Reina2009} from \ce{SiO2}/Si substrate\cite{VanDerZande2013}. After transferring, the PMMA film was removed by placing the samples in acetone for 12 hours. 

\newpage
\section{2. Laser irradiation parameters for phase transition}
To initiate the 2H-to-\Td\ phase transition of \ce{MoTe2}, we irradiated the encapsulated samples using continuous wave (CW) 532 nm laser and power of 21 mW at ambient conditions. The laser was focused by a 100× objective lens (N.A. = 0.9) and the resulting spot size on the substrate was around 1$\mu$m. The laser-irradiated area was patterned by rastering the laser spot with 200 nm point-to-point distance and 0.1 s exposure time per step. 

\section{3. S/TEM measurement}
\textit{In situ} TEM experiments were done in a Thermo Fisher Scientific Themis-Z aberration-corrected S/TEM operated at 80 kV. For atomic-resolution ADF-STEM imaging, the point resolution was about 1 \Ang\ with 25 mrad convergence semi angle, 35 pA probe current, 63 to 200 mrad collection semi angles, 20 pm pixel size and a total dwell time of 20 \dwell\ using 10-frame averages. For BFTEM, SAED, and DFTEM, the data were acquired with a Ceta 16M camera at parallel illumination using the three-condenser TEM mode. The electron dose rate was around 10\textsuperscript{3} \doserate\ and the exposure times for SAED and DFTEM were 2 to 5 s. Note that sparse dark pits were observed in DFTEM at the end of the \textit{in situ} imaging after an accumulated total dose around $4\times 10^{5}$ \dose. While the graphene encapsulation was unlikely to form holes under this condition, these dark pits were likely to be crystallographic defects such as voids formed by displacing atoms of the \ce{MoTe2} flakes. 

\newpage
\section{4. \textit{Ex situ} \Td-to-2H phase transition with annealing}
The \textit{ex situ} \Td-to-2H phase transition of \ce{MoTe2} (Figure 5 of the main text) was performed by an annealing process in a vacuum furnace. We annealed the laser-irradiated sample in a vacuum chamber (10\textsuperscript{-4} Torr) and slowly ramped up to targeted temperatures in 3 hours and held for another 3 hours. The targeted temperatures were set from 300 to 800 \degC. The furnace was naturally cooled to room temperature.

\section{5. Raman spectroscopy measurement}
\begin{figure}[H]
  \includegraphics[width=16cm]{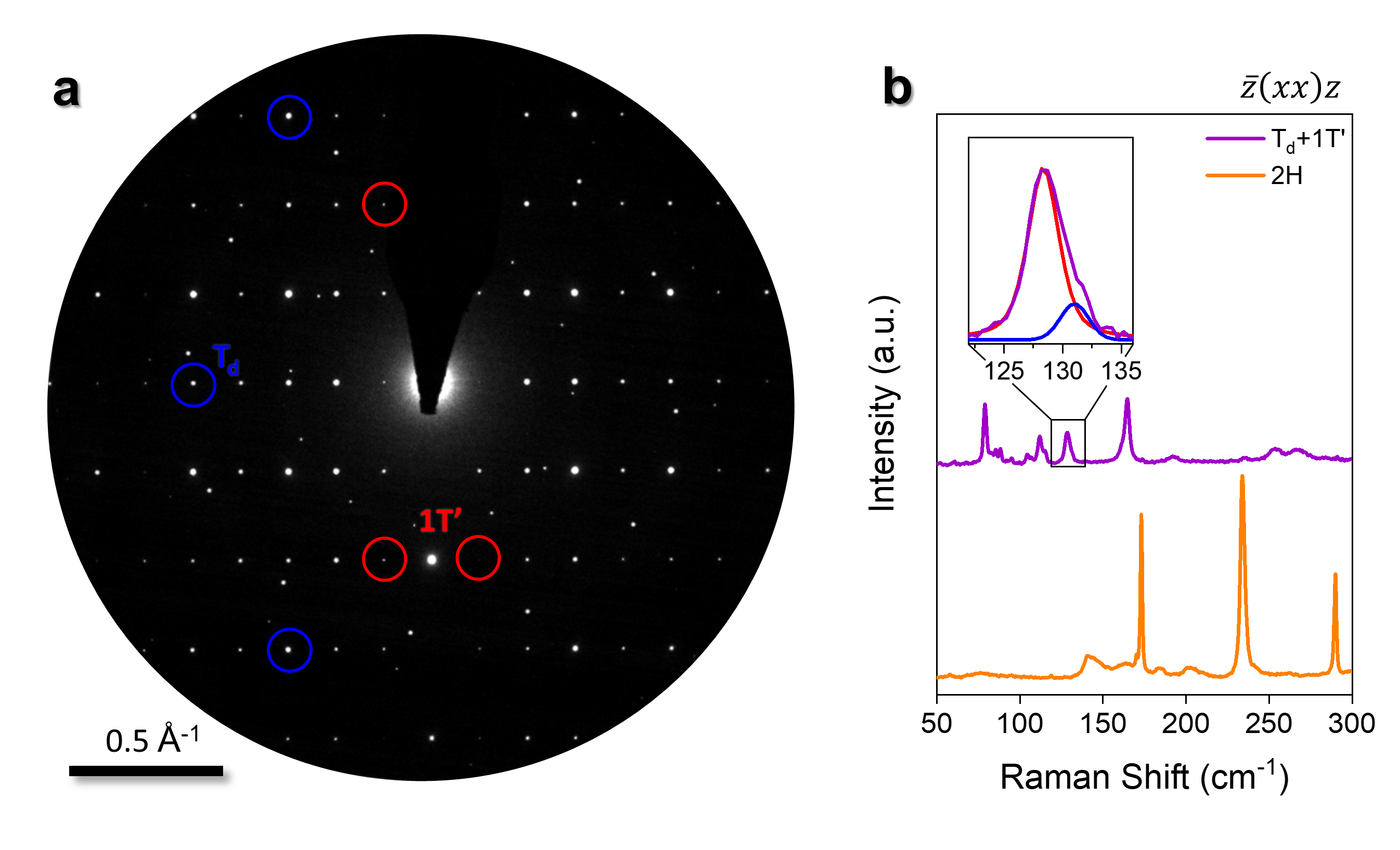}
  \caption{Mixture of 1T' and \Td\ phase characterized by TEM diffraction and polarized Raman spectroscopy. (a) Selected area electron diffraction pattern (SAED) acquired at a region with mixed 1T' and \Td\ phases. The red circles are Bragg peaks of \Td\ phase that would be absent if it were 1T' phase, however, the intensities are too weak for the region to be pure \Td\ phase, indicating a mixture of 1T' and \Td\ phases. The Bragg peaks inside the blue circles are also characteristic peaks of \Td\ phase that are absent in the 1T' phase. (b) Polarized Raman spectra of \Td\ + 1T'- (purple) and 2H-\ce{MoTe2} (orange). The 1T' and \Td\ phase are typically characterized by the peak splitting around 128 cm\textsuperscript{-1}: those with a split peak were identified as the \Td\ phase, and those with a single peak as the 1T' phase\cite{Cheon2021}. The blue peak in the inset indicates the presence of \Td\ phase. However, considering the SAED result in (a), our specimen shows a spatial inhomogeneity of mixture of 1T' and \Td\ phases.} 
  \label{fgr:S2}
\end{figure}

The linearly polarized Raman measurements (SI Figure \ref{fgr:S2}) were carried out in the backscattering geometry using 514.5 nm laser excitation. The input laser beam was focused onto the samples by a 50× microscope objective lens (0.8 NA), and the scattered light was collected and collimated by the same objective lens. To access the low-frequency range below 50 cm\textsuperscript{–1}, volume holographic filters (OptiGrate) were used to clean the laser lines and reject the Rayleigh-scattered light. A laser with a low power of 300 $\mu$W was used to avoid local heating. The Raman scattering signals were dispersed by a Jobin–Yvon iHR550 spectrometer with a 2400 grooves/mm grating (400 nm blaze) and detected by a liquid-nitrogen-cooled, back-illuminated CCD detector. An achromatic half-wave plate was used to rotate the polarization of the linearly polarized laser beam to the desired direction. The analyzer angle was set such that photons with polarization parallel to the incident polarization passed through. Another achromatic half-wave plate was placed in front of the spectrometer to keep the polarization direction of the signal entering the spectrometer constant with respect to the groove direction of the grating. The Raman spectra (Figure 5 of the main text) were acquired using a HORIBA LabRAM HR Evolution with the laser wavelength at 532 nm. To minimize the irradiation damage, the laser power was set below 5 mW with an acquisition time of 60 s. All measurements were conducted at ambient conditions.

\newpage

\newpage

\newpage
\begin{figure}[H]
  \includegraphics[width=16cm]{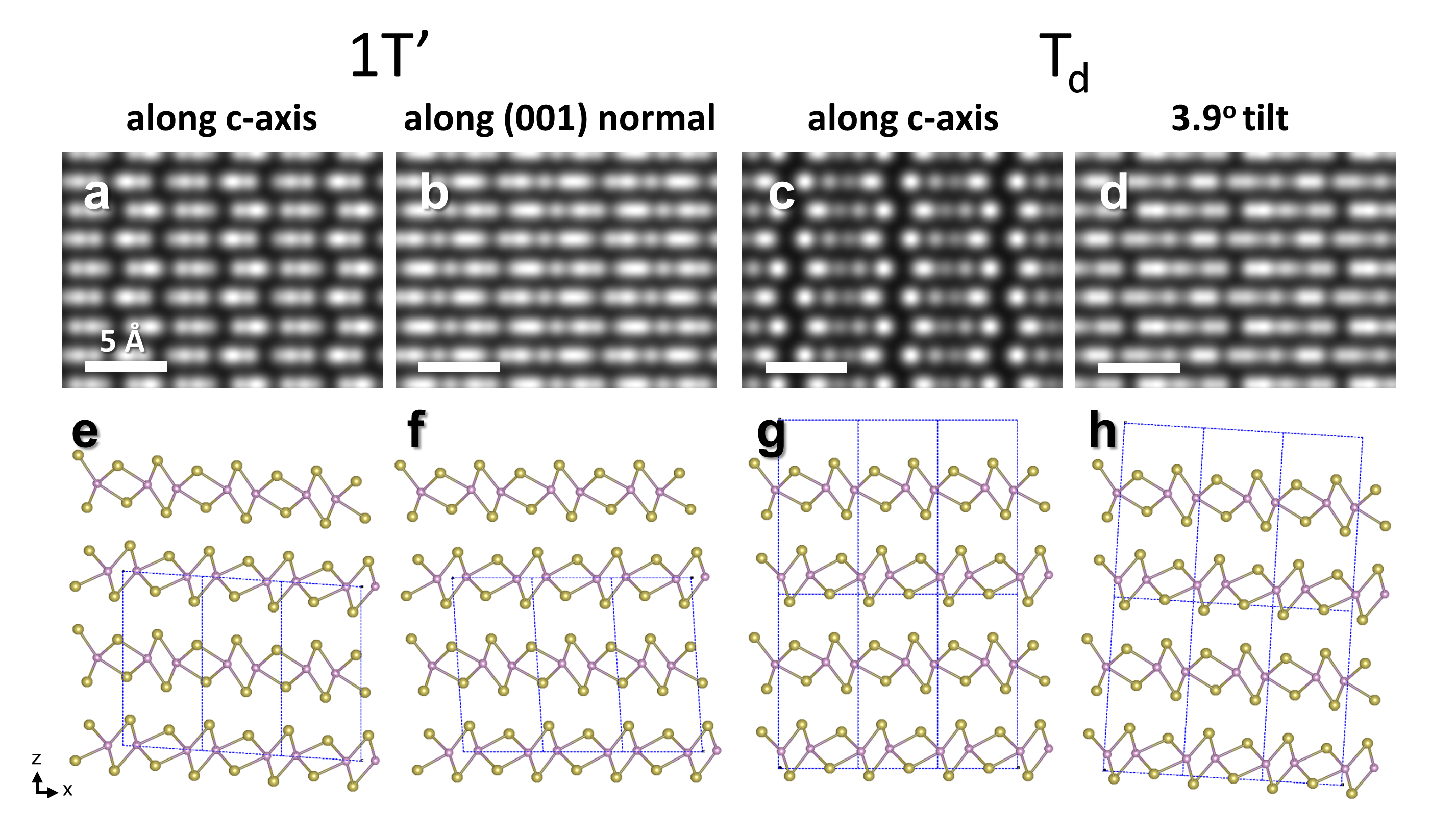}
  \caption{ADF-STEM image simulation of 1T'- and \Td-\ce{MoTe2}. (a--d) Simulated ADF-STEM images of 1T' and \Td\ phase at different orientations using semi-quantitative image simulation package\cite{Kirkland2013a} (e--h) Atomic models of 1T' and \Td\ phase at different orientations. The $3.9^{\circ}$ tilt angle is chosen to match the $\beta$ angle of 1T' phase.} 
  \label{fgr:S3}
\end{figure}

\newpage
\begin{figure}[H]
  \includegraphics[width=16cm]{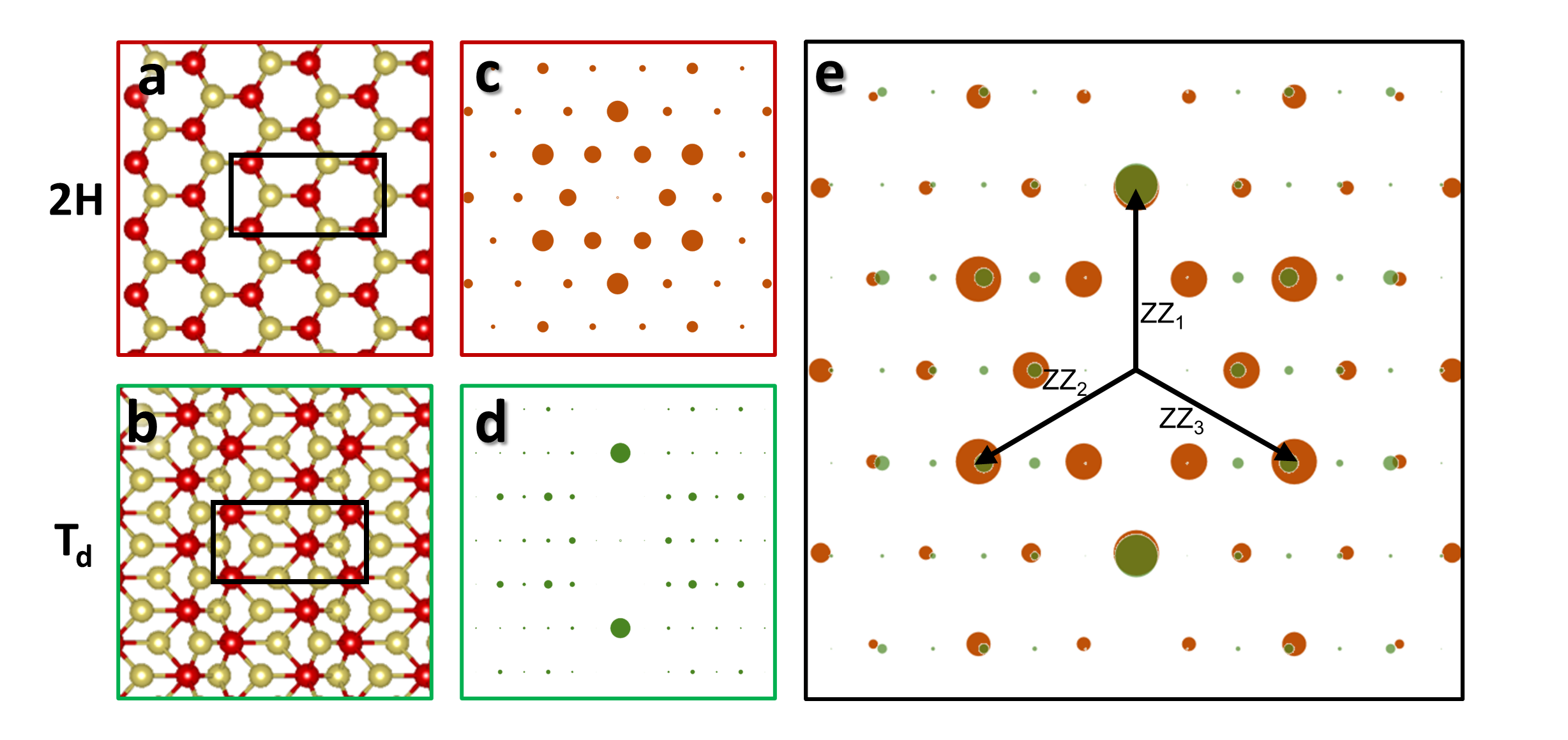}
  \caption{Crystallographic relation between the \Td\ and the 2H matrix. (a,b) Atomic models of top-view, monolayer 2H and \Td\ phases. (c,d) Simulated diffraction patterns of 2H and \Td\ phase. (e) Overlay of the simulated diffraction patterns of 2H and \Td\ phase. The \Td\ phase can be derived by shifting the chalcogen layers in 2H phase along one of the three arm-chair directions followed by some metal atom dimerization. Therefore, the b-axis of the derived \Td\ variants are parallel to one of the three zig-zag directions of the 2H matrix.} 
  \label{fgr:S4}
\end{figure}

\newpage
\begin{figure}[H]
  \includegraphics[width=14cm]{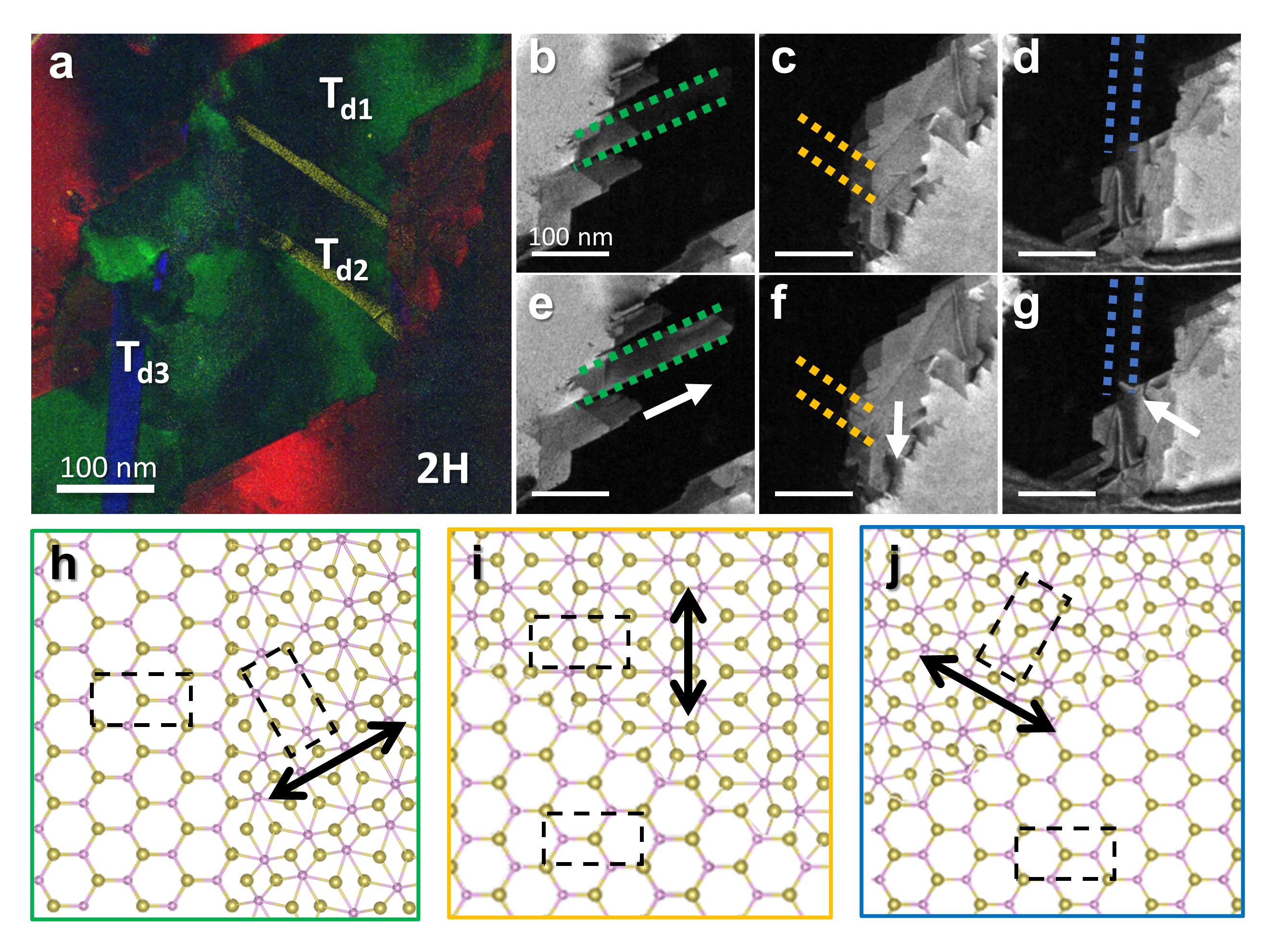}
  \caption{Anisotropic phase transition for all 3 \Td\ orientations. (a) Overlay of false-colored DFTEM images of the 2H matrix (red) and three different orientations of \Td\ grains. Reproduced from Figure 2f of the main text. (b--d) DFTEM images of 2H-\Td\ interfaces with different \Td\ phase orientations. (e--g) DFTEM images of 2H-\Td\ interfaces after heat pulses. The white arrows indicate the growth directions of the 2H phase front. The growth directions are parallel to the b-axis directions of the nearby \Td\ grains. (h--j) Atomic models of 2H-\Td\ interface with three different orientations. The b-axis directions of the \Td\ phases are marked by the black arrows.} 
  \label{fgr:S5}
\end{figure}

\newpage
\begin{figure}[H]
  \includegraphics[width=12cm]{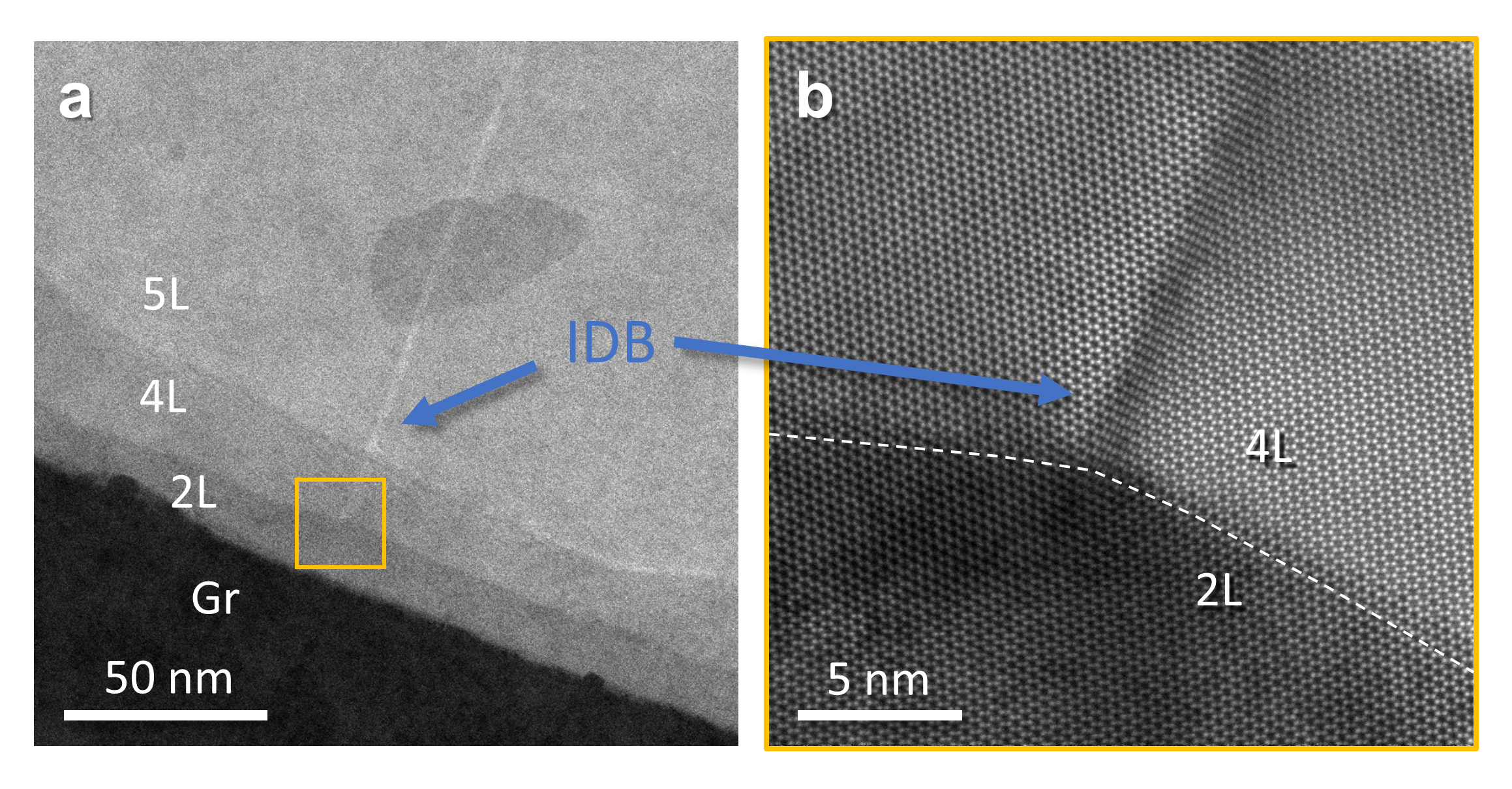}
  \caption{ADF-STEM images of an inversion domain boundary (IDB) at (a) lower and (b) higher magnifications. The field-of-view is marked by the yellow square. The IDBs are marked by the blue arrows. In this specific region, the IDB is only observed at the 4-layer region, indicating the IDB does not go all-the-way-through this 5-layer sample and suggesting an independent layer-by-layer phase transition mechanism. The layer number is determined by quantitative ADF-STEM intensities.} 
  \label{fgr:S6}
\end{figure}

\newpage
\begin{figure}[H]
  \includegraphics[width=12cm]{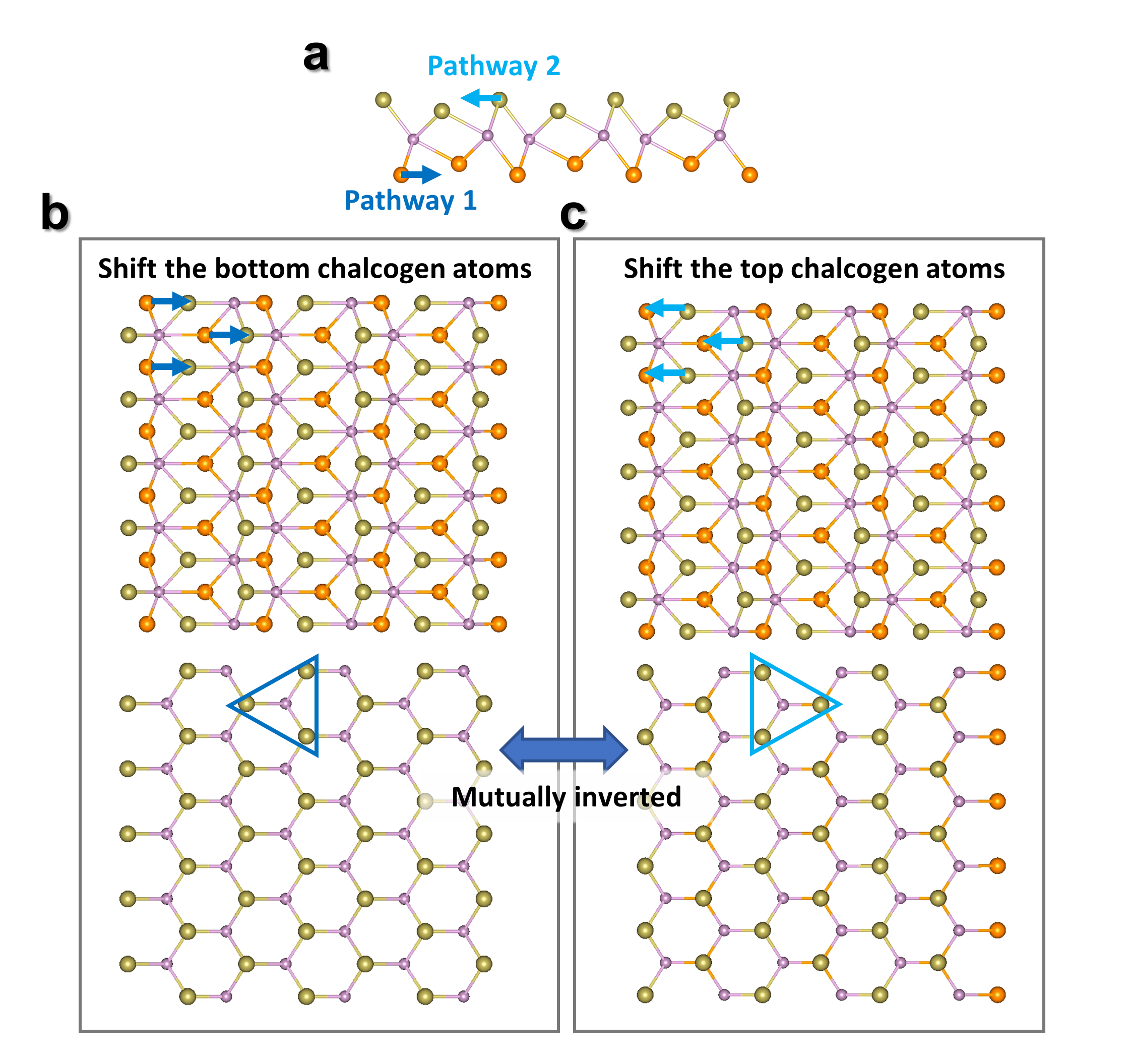}
  \caption{Formation mechanism of the inversion domain boundary. Atomic models of two potential pathways of \Td-to-2H phase transition in (a) side-view and (b,c) top-views. By sliding either the top or bottom chalcogen layers along the a-axis direction, 2H grains with opposite orientations can be derived from a single crystalline \Td\ grain. Therefore, shifting opposite layers of the chalcogen atoms in a \Td\ grain will generate an inversion domain boundary. The dark (Pathway 1) and light (Pathway 2) blue arrows indicate the sliding directions of each chalcogen layer.} 
  \label{fgr:S7}
\end{figure}

\makeatletter 
\setcounter{figure}{0}
\renewcommand{\figurename}{Movie}
\makeatother

\newpage
\begin{figure}[H]
  \includegraphics[width=10cm]{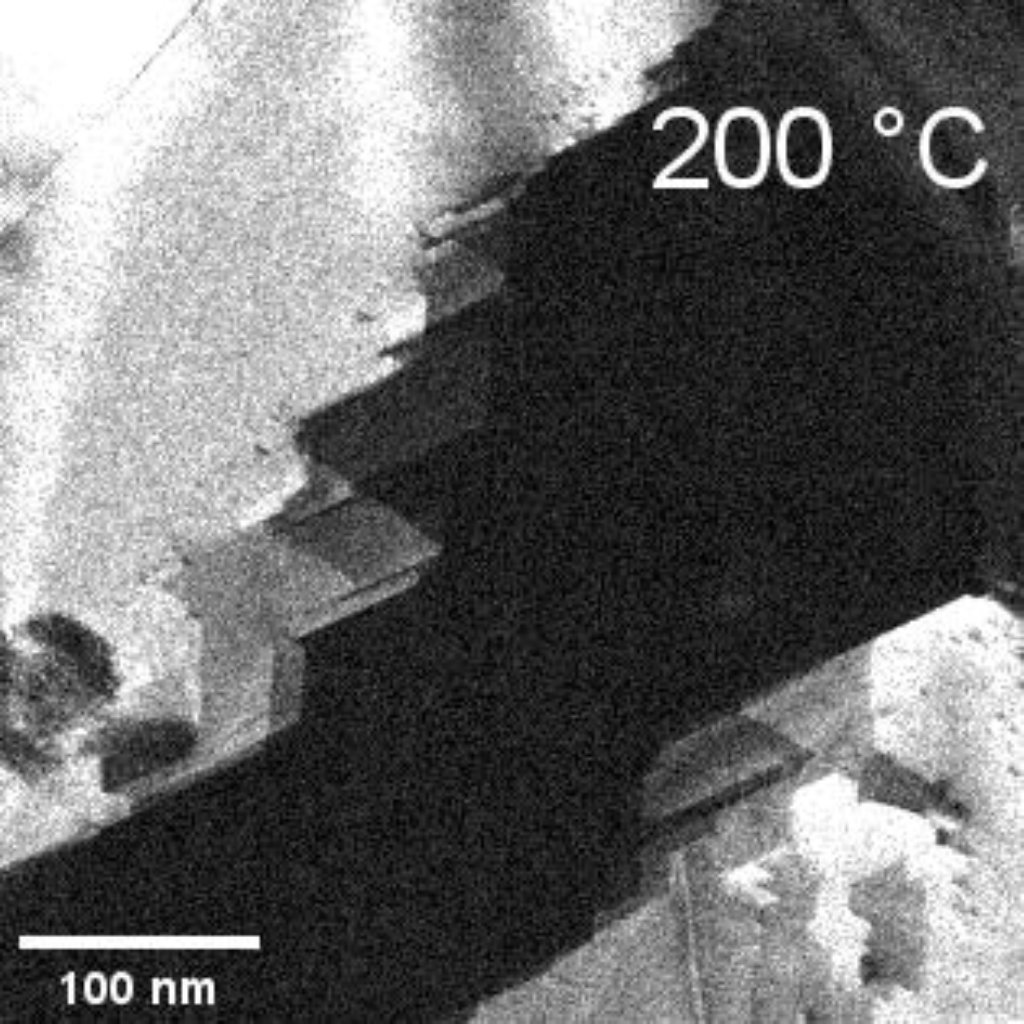}
  \caption{DFTEM video acquired using the $(\bar{1}100)$\textsubscript{2H} spot at room temperature after each heat pulse from 200 to 275 \degC, showing the in-plane, layer-by-layer, and anisotropic \Td-to-2H phase transition. The heat pulses range from 0.5 s to 1 min.} 
  \label{mov:S1}
\end{figure}

\newpage
\begin{figure}[H]
  \includegraphics[width=10cm]{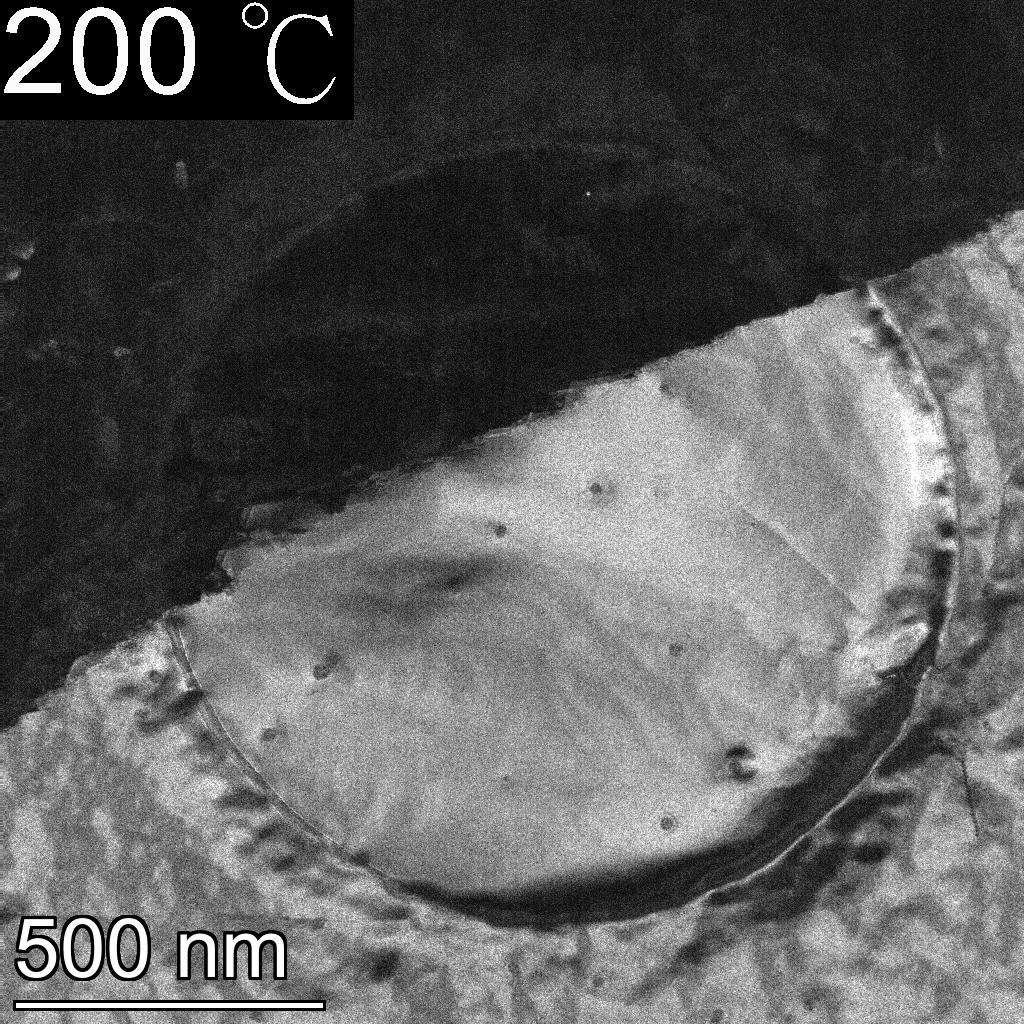}
  \caption{DFTEM video acquired using the $(\bar{1}100)$\textsubscript{2H} spot at room temperature after each heat pulse from 200 to 700 \degC, showing the anisotropic \Td-to-2H phase transition at lower temperatures. The phase transition then become more isotropic at temperatures above 400 \degC. We applied two rounds of heating, the first round is 200--400 \degC, while the second round is 200--700 \degC. The temperature intervals are all 25 \degC\ and the heat pulses are all 0.5 s.} 
  \label{mov:S2}
\end{figure}

\bibliography{My-Collection}